\documentclass[preprint,prb,superscriptaddress,longbibliography]{revtex4-2}
\usepackage[utf8]{inputenc}
\usepackage{graphicx}
\usepackage[colorlinks=true,citecolor=blue]{hyperref}
\usepackage{amsmath}
\usepackage{amssymb}
\usepackage{soul}
\usepackage{gensymb}
\usepackage{amsfonts,amssymb,amsbsy}
\usepackage{siunitx}
\usepackage{physics}
\usepackage{booktabs}
\usepackage[percent]{overpic}

\newcommand{\dIdV}{d$I$/d$V$~}
\newcommand{\ddIdV}{d$^2I$/d$V^2$~}
\usepackage{xcolor}
\usepackage{natbib}

\usepackage{lineno}

\begin{document}

\title{Observation of electromagnons in a monolayer multiferroic}

\author{Mohammad Amini}
\thanks{These authors contributed equally.}
\affiliation{Aalto University, Department of Applied Physics, 00076 Aalto, Finland}

\author{Tiago V. C. Antão}
\thanks{These authors contributed equally.}
\affiliation{Aalto University, Department of Applied Physics, 00076 Aalto, Finland}

\author{Liwei Jing }
\affiliation{Department of Physics, Nanoscience Center, 
University of Jyväskyl\"a, FI-40014 University of Jyväskyl\"a, Finland}

\author{Ziying Wang }
\affiliation{Aalto University, Department of Applied Physics, 00076 Aalto, Finland}

\author{Antti Karjasilta }
\affiliation{Aalto University, Department of Applied Physics, 00076 Aalto, Finland}

\author{Robert Drost}
\affiliation{Aalto University, Department of Applied Physics, 00076 Aalto, Finland}

\author{Shawulienu Kezilebieke}
\affiliation{Department of Physics, Department of Chemistry and Nanoscience Center, 
University of Jyväskyl\"a, FI-40014 University of Jyväskyl\"a, Finland}

\author{Jose L. Lado}
\email{Corresponding authors. Email: jose.lado@aalto.fi, adolfo.oterofumega@aalto.fi, peter.liljeroth@aalto.fi}
\affiliation{Aalto University, Department of Applied Physics, 00076 Aalto, Finland}

\author{Adolfo O. Fumega}
\email{Corresponding authors. Email: jose.lado@aalto.fi, adolfo.oterofumega@aalto.fi, peter.liljeroth@aalto.fi}
\affiliation{Aalto University, Department of Applied Physics, 00076 Aalto, Finland}

\author{Peter Liljeroth}
\email{Corresponding authors. Email: jose.lado@aalto.fi, adolfo.oterofumega@aalto.fi, peter.liljeroth@aalto.fi}
\affiliation{Aalto University, Department of Applied Physics, 00076 Aalto, Finland}

\begin{abstract}

Van der Waals multiferroics have emerged as a promising platform to explore novel magnetoelectric phenomena. Recently, it has been shown that monolayer NiI$_2$ hosts robust type-II multiferroicity, a giant dynamical magnetoelectric coupling at terahertz frequencies, and an electrically switchable spin polarization. These developments present the possibility of engineering ultrafast, low-energy-consumption, and electrically-tunable spintronic devices based on the collective excitations of the multiferroic order, electromagnons. However, their direct visualization in real space and in the monolayer limit remains elusive. Here, we report the atomic-scale observation of electromagnons in monolayer NiI$_2$ using low-temperature scanning tunneling microscopy by resolving the coherent in-gap excitations of the symmetry-broken multiferroic state. Comparison with first-principles, spin models, and Monte Carlo calculations reveals that the low-energy modes originate from electromagnon excitations, whose spectrum is gapped by the magnetoelectric coupling between the spins and the emergent electric polarization. Spatially resolved inelastic tunneling spectroscopy maps show a stripe-like modulation with the period of the spin spiral at electromagnon energies, consistent with spin-orbit-assisted inelastic tunneling. These results provide direct evidence of the real-space structure of electromagnons and provides a methodology to probe these modes at the atomic scale, opening avenues for electrically tunable spintronics.

\end{abstract}

\date{\today}
\maketitle 
\newpage

\section*{Introduction}
Quantum materials provide a unique playground for the emergence of a variety of exotic excitations, ranging from fermionic and bosonic to anyonic quasiparticles. A paradigmatic family of quasiparticles arises from spontaneous symmetry breaking, which results in bosonic excitations associated with the lifting of a continuous symmetry.
Well-known examples of these bosonic modes are phonons in crystal lattices and magnons in magnetic systems,\cite{Venema2016} where these emergent bosonic modes critically influence the dynamic properties and transport phenomena.\cite{Maldovan2013,Chumak2015,Pirro2021}
Among the materials with symmetry-broken phases, type-II magnetoelectric multiferroics hold significant technological potential due to the simultaneous presence of magnetic and electric orders and their strong magnetoelectric coupling.\cite{Fiebig2016,Spaldin2019} 
In these systems, electromagnons, collective excitations concurrently involving electric and magnetic degrees of freedom, emerge as bosonic modes that can strongly couple to electric and magnetic fields.\cite{Takahashi2011,PhysRevLett.111.037204,Matsubara2015} This unique feature offers compelling opportunities for the development of ultrafast, electrically-tunable and low-energy consumption optical and spintronic devices.\cite{Kubacka2014,Masuda2021,Ogino2024}

Van der Waals multiferroics have emerged as a promising platform to explore novel magnetoelectric phenomena.\cite{Gao2021,Man2023,Tang2025} Recently, it has been shown that monolayer NiI$_2$ hosts robust type-II multiferroicity down to the two-dimensional limit \cite{Ju2021, Song2022, Fumega2022, Amini2024,Wang2024_SP,Miao2025,4hzc-bm2f}. NiI$_2$ is a prototypical type-II spin-spiral multiferroic,\cite{Billerey1977,Kuindersma1981,Kurumaji2013,Ju2021,Katsura2005,Mostovoy2006,Jiangping2008,Fumega2022,Tseng2025} where the interplay between spin-spiral magnetic order and strong spin-orbit coupling, inherent to iodine atoms, generates simultaneous magnetic and ferroelectric orders.\cite{Amoroso_2020,Fumega2022,Li2023_realisticspinmodelnii2,Sdequist2023} 
The strong magnetoelectric coupling associated with this type of multiferroicity has been widely demonstrated for NiI$_2$.\cite{Kurumaji2013,Son2022,Amini2024,Wu2024,Song2025}
Remarkably, Gao et al. have recently reported a giant dynamical magnetoelectric coupling in NiI$_2$ using time-resolved second-harmonic generation and reflective Kerr rotation. \cite{Gao2024}
Their findings highlight pronounced optical activity at terahertz frequencies linked directly to resonant electromagnon excitations.\cite{Gao2024,Kim_THzNiI2_PhysRevB.108.064414} These findings make single layer NiI$_2$ particularly promising for integration into van der Waals heterostructures and moiré architectures as a route to induce magnetoelectric tunability in complex systems.\cite{vdwHT2013,Andrei2021,Song2021_twistedcri3,Xu2021,Antao2024}

Direct atomic-scale visualization of electromagnon modes in monolayer NiI$_2$ has remained elusive, limiting deeper exploration and development of magnetoelectric spintronics in complex van der Waals-based systems.
In this work, we overcome this fundamental limitation through a combination of low-temperature scanning tunneling microscopy (STM) experiments, theoretical ab-initio techniques and spin models, enabling the direct atomic-scale visualization of electromagnons in monolayer NiI$_2$. Temperature-dependent STM imaging is used to follow the thermal melting of the multiferroic stripes, and low-temperature inelastic tunneling spectroscopy (IETS) to identify and distinguish the electromagnon and phonon modes. Our theoretical analysis, incorporating first-principles calculations, spin-model and Monte Carlo simulations, shows that the magnetoelectric coupling between the magnetic excitations and the emergent electric polarization opens a gap in the electromagnon spectrum, which we resolve in the inelastic spectra. Spatially resolved IETS mapping reveals a modulation with the period of the spin spiral at electromagnon energies, providing an additional, real-space check of the electromagnon character of these excitations and of their internal atomic-scale structure. These findings provide a methodology for probing and characterizing electromagnon excitations in monolayer and other low-dimensional multiferroic systems, thereby paving the way for innovative device concepts to engineer ultrafast, low-energy-consumption, and electrically-tunable spintronic devices based on the collective excitations of the multiferroic order \cite{Takahashi2011,PhysRevLett.111.037204,Matsubara2015,Gao2024,Kim_THzNiI2_PhysRevB.108.064414}. 

\begin{figure*}[ht!]
    \centering
    \includegraphics[width =1 \textwidth]{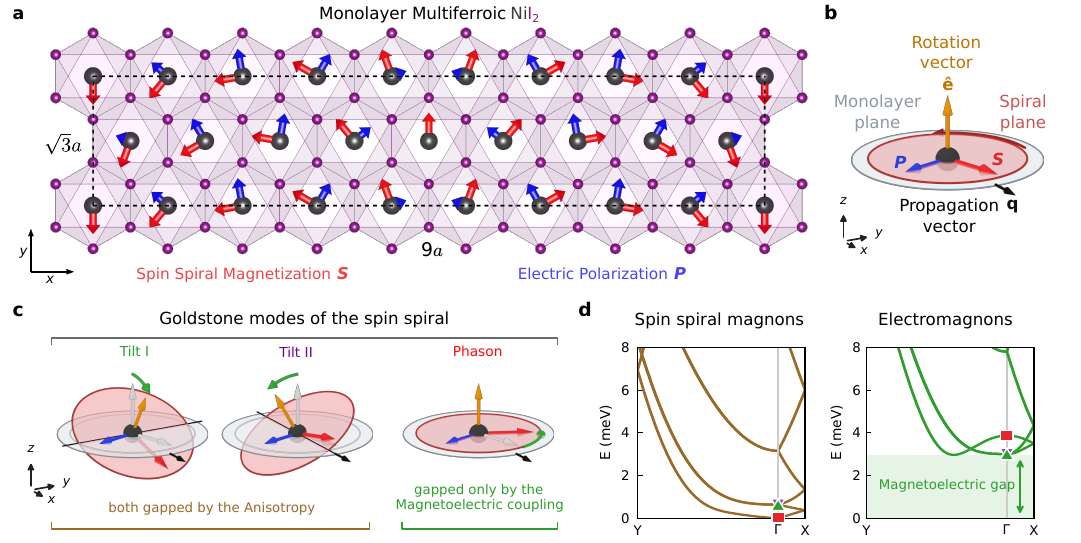}
    \caption{Multiferroic order and collective modes in monolayer multiferroic NiI$_2$. a) Schematic of monolayer multiferroic NiI$_2$. The spin-spiral magnetic order and the electric polarization, with half the periodicity of the magnetic order, are depicted as red and blue arrows, respectively. The dashed rectangle marks the $9a\times\sqrt{3}a$ supercell. b) Schematic of a single spin of the spiral. The local magnetization ($\textbf{S}_i$) rotates in the spiral plane, about the rotation vector $\hat{\textbf{e}}$, as it propagates along $\textbf{q}$, and the emergent electric polarization ($\textbf{P}_i$) is perpendicular to $\textbf{S}_i$.
    c) The three Goldstone modes of the spin spiral, given by two tilts of the spiral plane, about an in-plane axis perpendicular to $\textbf{q}$ (tilt I) and about $\textbf{q}$ (tilt II), both gapped by the magnetic anisotropy, and the phason, a sliding of $\textbf{S}_i$ within the spiral plane, gapped only by the magnetoelectric coupling.
    d) Magnon bands of the spin model without (spin-spiral magnons) and with the magnetoelectric coupling $D$ (electromagnons). The symbols at $\Gamma$ mark the three Goldstone modes of c. The magnetoelectric coupling opens a gap in the electromagnon spectrum. }
    \label{Figure1}
\end{figure*}

\section*{Results}

\subsection*{Theoretical insight into multiferroicity and electromagnon modes}

To understand the nature of electromagnon excitations in monolayer NiI$_2$, we begin by examining the theoretical framework underlying its multiferroicity. A $J_1$–$J_3$ Heisenberg spin model with single-ion anisotropy captures the key magnetic features of this two-dimensional (2D) system. At low temperatures, monolayer NiI$_2$ exhibits a spin-spiral magnetic order, driven by competing ferromagnetic nearest-neighbor ($J_1$) and antiferromagnetic third-neighbor ($J_3$) exchange interactions (Fig.~\ref{Figure1}a). The anisotropy term stabilizes the rotation plane of the spin spiral.\cite{Sdequist2023} Through the inverse Dzyaloshinskii-Moriya interaction (DMI),\cite{Katsura2005,Mostovoy2006,Sergienko2006} the combination of the spiral magnetic order and the strong spin-orbit coupling $\lambda$ of iodine atoms gives rise to an emergent electric polarization $\textbf{P}_i=\lambda\textbf{S}_i\times (\nabla \times \textbf{S}_i)$ perpendicular to the spin $\textbf{S}_i$ at each site $i$. The emergent polarization exhibits half the periodicity of the spin spiral (Fig.~\ref{Figure1}a). More general forms of the magnetoelectric response, allowed by the lower symmetry of the triangular lattice,\cite{Kaplan2011,Yu2025} change the direction and strength of the static polarization, which enter our model only through the effective coupling introduced below. As a result of the crystalline and multiferroic orders present in monolayer NiI$_2$, two dominant low-energy collective excitations emerge, phonons and electromagnons. Using first-principles calculations combined with spin models (See Methods), we compute the energy spectrum of these excitations and distinguish their origins.
The phonon mode that couples most strongly to the tunneling electrons, an out-of-plane vibration of the layer, lies at about 16 meV.
In contrast, the low-energy modes, from about 3 meV upwards, are primarily associated with electromagnons, collective oscillations arising from the coupled dynamics of the spin and polarization (Fig.~\ref{Figure1}b,c).
The spin spiral breaks the continuous spin-rotation symmetry and hosts three Goldstone modes (Fig.~\ref{Figure1}c), namely two tilts of the spiral plane, about an in-plane axis perpendicular to $\textbf{q}$ (tilt I) and about $\textbf{q}$ itself (tilt II), and the phason, in which the spins slide within the spiral plane and shift the spiral and the polarization stripes along $\textbf{q}$. The single-ion anisotropy fixes the orientation of the spiral plane and therefore gaps the two tilt modes, but it leaves the phason gapless (spin-spiral magnons in Fig.~\ref{Figure1}d).
Crucially, the emergent polarization introduces an energy cost in the spin model for the spin spiral associated with fluctuations of the electric polarization (See Methods) intrinsic to the multiferroic phase. We describe it with a magnetoelectric coupling $D\sum_i(\textbf{S}_i\cdot\hat{\textbf{P}}^0_i)^2$, where $\hat{\textbf{P}}^0_i$ is the direction of the static polarization at site $i$. Since $\textbf{S}_i\perp\textbf{P}_i$, this term has no effect on the ground-state spin configuration, but it penalizes the precession of the spins towards the local polarization. It therefore gaps the phason and raises the tilt modes, opening a gap in the whole electromagnon spectrum (electromagnons in Fig.~\ref{Figure1}d).
The magnetoelectric gap is absent in the magnon spectrum of purely magnetic spin-spiral systems, in which the phason remains a gapless Goldstone mode. Therefore, the gapped spectrum constitutes a distinctive fingerprint of electromagnons. Spatially-resolved IETS  \cite{Ternes2015,Spinelli2014} provides a route to experimentally resolve this gap and to probe the real-space structure of the electromagnon excitations at the atomic scale.

\begin{figure*}[tp!]
    \centering
    \includegraphics[width =1 \textwidth]{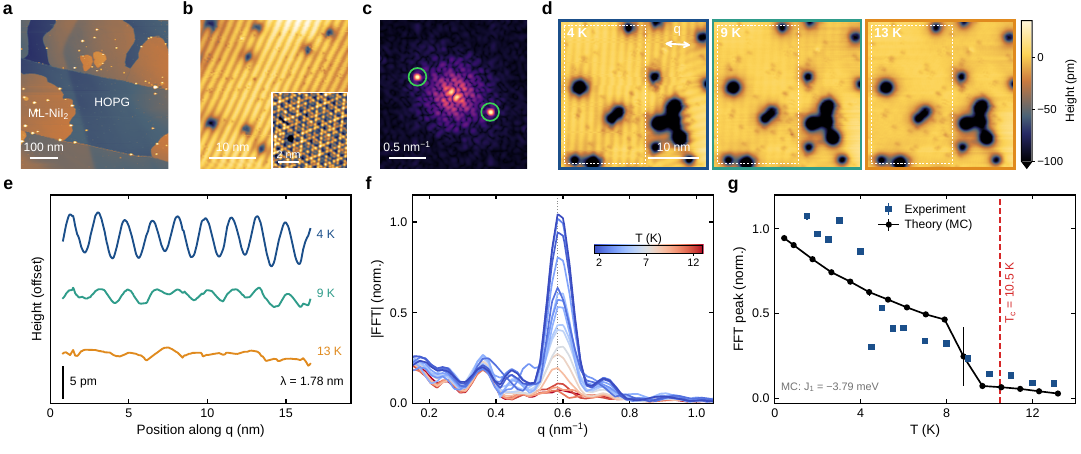}
    \caption{Temperature-dependence of the multiferroic order. a) Large area STM scan of monolayer NiI$_2$ on HOPG (image size $ 495\times 495 \mathrm{~nm}^{2}, V= 1.2 \mathrm{~V}, I= 3.6 \mathrm{~pA} $). b) STM scan of multiferroic stripes in monolayer NiI$_2$ (image size $ 30\times 30 \mathrm{~nm}^{2}, V= 1 \mathrm{~V}, I= 25 \mathrm{~pA} $). Inset is a zoomed-in atomic resolution scan of the same island ($ 7\times 7 \mathrm{~nm}^{2}, V=-1 \mathrm{~V}, I=300 \mathrm{~pA} $). c) The corresponding FFT of the data in panel b. The green circles indicate the peaks associated with the stripe modulation.
    d) STM scans of the same area at 4, 9 and 13~K. Increasing the temperature causes the multiferroic stripes to disappear (image size $ 29.2\times 29.2 \mathrm{~nm}^{2}, V= 0.7 \mathrm{~V}, I= 10 \mathrm{~pA} $). The arrow indicates the direction of $q$, parallel to the stripe wavevector $k_P=2q$. e) Height profiles along $q$, averaged inside the dashed boxes in d.
    f) FFT line cuts along $q$ for all temperatures between 1.5 and 13~K.
    g) Evolution of the stripe amplitude, given by the FFT peak height normalized to its low-temperature value ($A/A_0$), as a function of temperature (squares), compared with the stripe amplitude $S_P(2q)$ obtained from Monte Carlo (MC) simulations of the spin model with $J_1=-3.79$~meV (line, see SM Section S5). The dashed line marks $T_c=10.5$~K, the temperature at which the stripe order is lost.}
    \label{Figure2}
\end{figure*}

\subsection*{Temperature-dependent observation of multiferroicity}
Guided by these theoretical insights, we proceed with the experimental investigation of the multiferroic order and its collective modes in monolayer NiI$_2$. We conducted low-temperature STM experiments on monolayer NiI$_2$ grown by molecular beam epitaxy (MBE) on highly oriented pyrolytic graphite (HOPG) (See Methods). Large-area STM scans confirm the successful synthesis of monolayer NiI$_2$ islands and reveal a uniform topography across extended terraces (Fig.~\ref{Figure2}a). Spectroscopic measurements at high bias confirm the insulating character of the monolayer, with a band gap of approximately 2.3 eV (See Supplementary Materials, SM), establishing a suitable electronic scenario for exploring in-gap excitations associated with the symmetry-breaking orders. By obtaining atomic-resolution STM scans at bias voltages at the conduction band, STM imaging reveals a stripe modulation superimposed on the atomic lattice (Fig.~\ref{Figure2}b). Fourier analysis of this image allows for a direct determination of the multiferroic periodicities (Fig.~\ref{Figure2}c). 
The two peaks enclosed with green circles correspond to the electronic modulation associated with ferroelectric polarization. 
Considering that the ferroelectric order has half the periodicity of the magnetic spin spiral, we can directly determine the $q$ vector of the spin spiral as half of the wavevector given by the peaks encircled in green. This enables the derivation of the relationship between the magnetic exchange interactions $J_3/J_1\approx -0.3$, giving rise to the periodicity of the multiferroic order highlighted in Fig.~\ref{Figure1}a.
This initial characterization of monolayer NiI$_2$ on HOPG is consistent with the coexistence of electric and magnetic order at the atomic scale, a hallmark of spin-spiral multiferroicity as previously reported for this system.\cite{Amini2024,Miao2025,4hzc-bm2f} The response of the stripe domains to an out-of-plane magnetic field (SM Section S3) shows their coupling to the magnetic order, and their manipulation by the electric field of the tip was shown in our previous work.\cite{Amini2024}

Next, we analyze the thermal evolution of the multiferroic order by acquiring temperature-dependent STM images of the same area. As the temperature increases, the stripe modulation progressively weakens and ultimately vanishes at high temperature, indicating the thermal melting of the multiferroic stripes (Fig.~\ref{Figure2}d,e). Fourier transforms of each image confirm this loss of modulation, as evidenced by the disappearance of the peak associated with the stripe order in the FFT line cuts (Fig.~\ref{Figure2}f). We fit Gaussians to the FFT peaks of the stripes as a function of temperature to quantify their evolution and to extract the temperature at which the stripe order is lost ($T_c$). As the temperature increases, $A/A_0$ decreases and at temperatures around $T_c$, the peak amplitude reaches the experimental noise floor. 

Importantly, this temperature sets the energy scale for the magnetic exchange interactions responsible for the spin-spiral ground state, and thus also governs the dynamics of the associated low-energy excitations.
To extract this energy scale, we performed classical Monte Carlo (MC) simulations of the spin model in which the emergent polarization of every spin configuration is computed, so that the amplitude of the polarization stripes, given by the polarization structure factor at $2\textbf{q}$, $S_P(2\textbf{q})$, can be compared directly with the measured stripe amplitude (SM Section S5).
A comparison of the measured stripe amplitude with the MC calculation as a function of temperature (Fig.~\ref{Figure2}g) allows estimation of the energy scale of the magnetic exchange interactions, with first neighbor ferromagnetic $J_1=-3.79$ meV and third neighbor antiferromagnetic $J_3=1.15$ meV, with $J_3/J_1$ fixed by the stripe period. The same simulations give $T_c\approx10.6$ K from the width of the spin structure factor peak of the whole simulation box (SM Section S5). It is worth noting that the spin model of Eq.~\ref{eq:Spin_Hamiltonian}, which describes both the thermal fluctuations of the spin spiral and its polarization (SM Section S5), retains the continuous symmetry of spin rotations about the out-of-plane axis, which shift the phase of the spin spiral. In agreement with the Mermin-Wagner theorem \cite{Mermin1966}, the stripes can therefore only display quasi-long-range order in two dimensions. The melting of the stripes is consistent with a Kosterlitz-Thouless-type transition \cite{Kosterlitz1973}, whereas the uniform polarization survives up to a higher temperature in the simulations (SM Section S5), and $T_c$ thus denotes the temperature at which the stripe order is lost on the scale of our images. These results set the energy range for the electromagnon modes (Fig.~\ref{Figure1}d).

\subsection*{Observation of electromagnon excitations}

\begin{figure*}[h!]
    \centering
    \includegraphics[width =1 \textwidth]{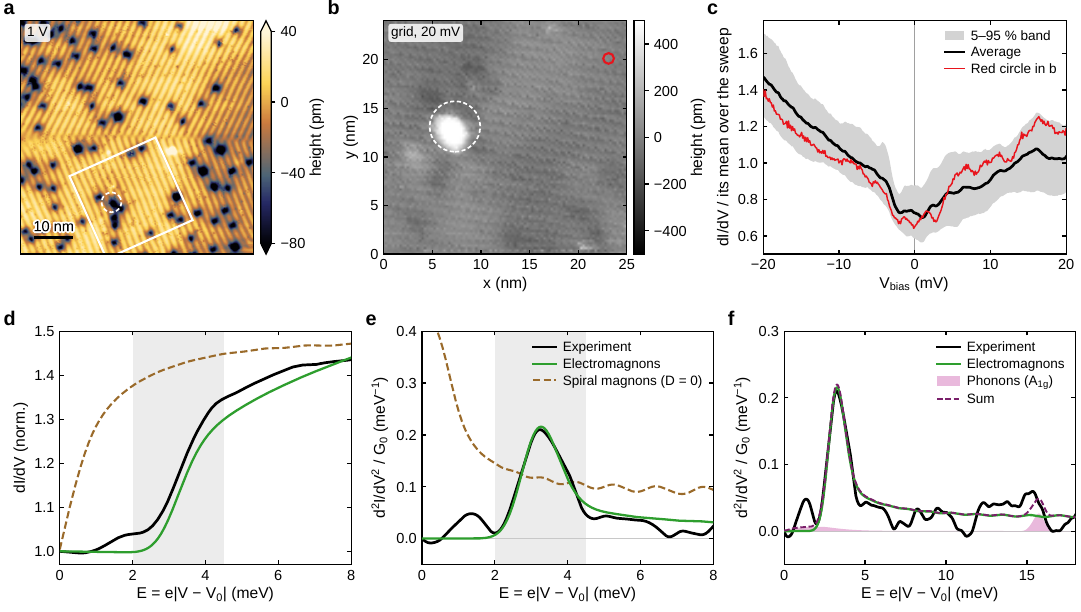}
    \caption{Low-energy excitations in monolayer multiferroic NiI$_2$. a) Topography of the region around the spectroscopic grid at high bias ($60\times60$~nm$^2$, $V=1$~V, $I=25$~pA, $T=3$~K), showing the ferroelectric stripes. The white rectangle marks the area of the grid of b, and the dashed circle the adsorbate of panel b. b) Height data from a spectroscopy grid experiment ($100\times96$ spectra over $25\times24$~nm$^2$, set point +20 mV, 0.5 nA). Adsorbate area removed from subsequent analysis is marked with a white dashed circle. c) All spectra outside the adsorbate area, divided by their mean over the sweep, shown as the 5–95\% band (gray), the average (black) and one example spectrum (red) acquired at the red circle in panel b. d) Statistical analysis of the low-temperature IETS, given by the average \dIdV of the mirror-symmetric group of spectra of the spectroscopic grid (1869 spectra, SM Section S7), symmetrized and normalized at $V_0$, as a function of $E=e|V-V_0|$ ($V_0=0.41$~mV) (black), compared with the spectra of the spin model with (green, electromagnons) and without (brown, spin-spiral magnons, $D=0$) the magnetoelectric coupling. e) Corresponding \ddIdV (black, symmetrized, with the $+V$ and $-V$ branches shown in SM Fig.~S8) and model spectra. The shaded region (2.0--4.5~meV) marks the fitting window. f) \ddIdV up to 18~meV compared with the electromagnon spectrum, the contribution of the out-of-plane A$_{1g}$ phonon band (pink) and their sum.
    }
    \label{Figure3}
\end{figure*}

Having characterized the multiferroic ground state and its temperature evolution, we now turn to the nature and spatial structure of the low-energy excitations associated with this phase. Low-temperature IETS offers a powerful probe to directly access these excitations with atomic-scale spatial resolution.
Figure~\ref{Figure3} shows the results of a high energy resolution \dIdV grid spectroscopy experiment ($100\times96$ spectra) at $T= 350$ mK.
At the lowest temperature, the spectra show bias-symmetric features, as expected for inelastic excitations; their evolution with temperature between 0.35 and 7.5~K is shown in SM Section S8 and Fig.~S12.
Figure~\ref{Figure3}a shows the region around the grid imaged at high bias, where the ferroelectric stripes are visible, and Fig.~\ref{Figure3}b,c show the height map recorded with the grid, and the average and an example single spectrum from the spectroscopy grid. To analyze these features statistically, we classified the \dIdV spectra into groups of similar spectra (SM Section S7). Figure~\ref{Figure3}d shows the average of the group with mirror-symmetric spectra. Based on the analysis, the inelastic \dIdV signal is flat around zero bias and rises in a step at $|V-V_0|\approx3$ mV, which appears in \ddIdV as a peak at 3.3 meV at both bias polarities (SM Fig.~S8). This is the gapped excitation spectrum expected for electromagnons (Fig.~\ref{Figure1}d). The local spin spectral function of the spin model (See Methods), with $J_1$ and $J_3$ fixed by the temperature dependence of Fig.~\ref{Figure2} and only the magnetoelectric coupling $D$, an overall intensity and the broadening fitted, reproduces both the onset and the peak for $D=3.0|J_1|=11.4$ meV (green), with a gap of 2.97 meV. In contrast, the spectrum of the spin-spiral magnons without magnetoelectric coupling ($D=0$, brown) is gapless and does not describe the data. 

We also observe weaker features in the spectra. A feature at about 1.3 meV inside the gap, more pronounced at positive bias, changes sign and bias polarity between different regions of the grid (SM Section S7 and Fig.~S8). Since an inelastic channel can only add conductance at both bias polarities, we attribute it to an elastic, spatially varying contribution to the local density of states. It is unlikely to correspond to phonons, as in addition to the bias asymmetry, the closest phonon modes of the multiferroic phase are expected to be at a higher energy (1.8 to 2.6 meV). At higher energies (Fig.~\ref{Figure3}f), some of the experimental features can correspond to the out-of-plane A$_{1g}$ phonon band (at 15.5 meV). This is a near-$\Gamma$ vibration that modulates the tip-layer distance and therefore is expected to couple to the tunneling electrons (See Methods). Including this phonon channel with a fitted weight improves the description of the spectrum above 5 meV, with phonons accounting for about 7\% of the inelastic weight. The remaining structure between 5 and 15 meV differs between the two bias polarities and is most likely elastic, although phonons may also contribute.

\subsection*{Spatial structure of the electromagnon excitations}

\begin{figure*}[!htbp]
    \centering
    \includegraphics[width =1 \textwidth]{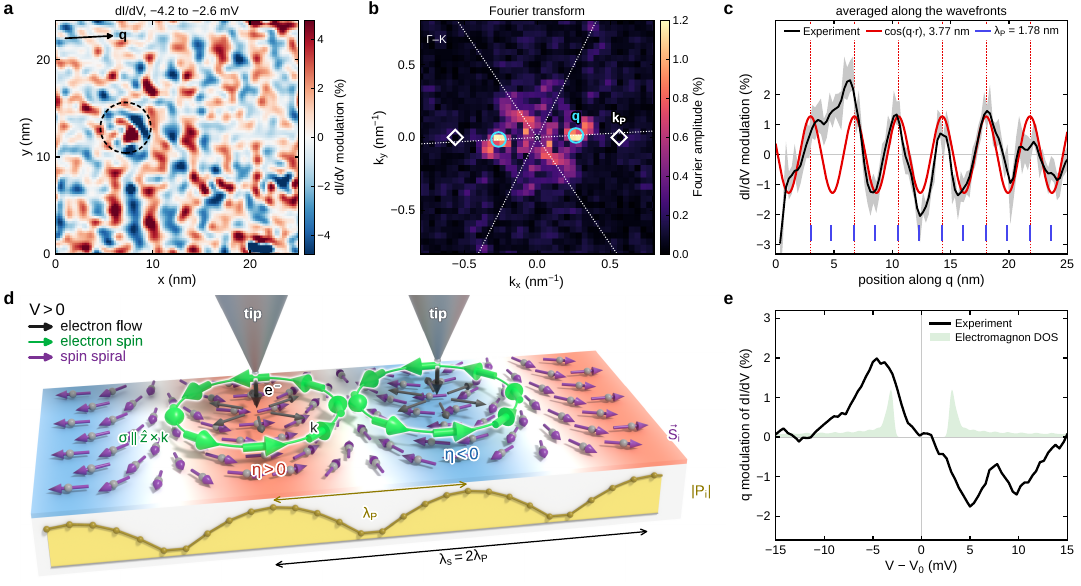}
    \caption{Electromagnon internal structure in monolayer multiferroic NiI$_2$. a) Normalized \dIdV map averaged between $-4.2$ and $-2.6$~mV ($V-V_0$), in the electromagnon region, extracted from a grid spectroscopy experiment described in Fig.~\ref{Figure3}. The smooth background contrast has been removed and the map is band-passed between 1.25 and 5.5~nm as described in SM Section S7. The dashed circle marks an adsorbate excluded from the analysis and the arrow indicates the stripe wavevector $q$. b) Fourier amplitude of the unfiltered map (amplitude of the cosine, in \% of the \dIdV signal). The circles mark $\pm q$, the diamonds the wavevector of the polarization stripes $k_P$ ($\lambda_P=1.78$~nm), and the dotted lines the $\Gamma$--K directions. c) Profile of the map averaged along the wavefronts of $q$ (grey, $\pm1$ standard error), with a cosine of period 3.77~nm (red). The red dashed lines mark its maxima and the blue ticks the period of the polarization stripes. Deviation from the q-cosine modulation can be observed around 5 nm, corresponding to the position of the adsorbate. d) Schematic of the spin-orbit-assisted tunneling, a local analog of the spin Hall effect, at $V>0$ (the same tip at two positions half a spin period apart). Electrons spreading from the tip with momentum $k$ acquire a spin $\sigma\parallel\hat{\textbf{z}}\times k$ (green) that winds around the tip, along the circulation of the spins (purple) at one position ($\eta>0$) and against it at the other ($\eta<0$). The contrast $\eta$ (top face) thus has the spin-spiral period $\lambda_s$ and reverses at $V<0$, while $|\textbf{P}_i|$ (front face, yellow) has half this period. e) Measured $q$ modulation of \dIdV as a function of bias in the mirror-symmetric group of spectra (black), projected on the stripe phase of $\ln[G(-V)/G(+V)]$, with the electromagnon density of states $S(|E|)$ of Fig.~\ref{Figure3} (green shading, arbitrary units).
    }
    \label{Figure4}
\end{figure*}

Finally, beyond identifying the spectral fingerprints of electromagnons, we exploit the spatial resolution of IETS to directly visualize the internal structure of these modes. Fig.~\ref{Figure4}a shows the \dIdV map at the electromagnon peak, averaged between $-4.2$ and $-2.6$ mV (data processed as described in detail in SM Section S7). A stripe-like modulation is visible across the whole frame. Its Fourier transform shows a single pair of peaks at $\pm q$ and no peak at the wavevector of the polarization stripes $k_P$ (Fig.~\ref{Figure4}b), and this peak lies far above the level of maps with randomly shuffled pixels (SM Section S7). The other features in the FFT are weaker than the peaks at $\pm\textbf{q}$, and we do not assign them. The profile averaged along the wavefronts (Fig.~\ref{Figure4}c) gives a period of $3.77\pm0.13$ nm and an amplitude of 1.3\% of the \dIdV signal. This period corresponds to twice that of the ferroelectric polarization modulation measured in the same area ($2.12\lambda_P$, with $\lambda_P=1.78$ nm), that is, to the period of the spin spiral, and the stripes run along the same $\Gamma$--K direction. We note that the spin period found in our experiment, about $9a$, is different from the period of about $4.5a$ found with spin-polarized STM in monolayer NiI$_2$.\cite{Wang2024_SP,Miao2025,4hzc-bm2f} The modulation is present at both bias polarities throughout the electromagnon energy range, and it is weaker at higher bias (SM Fig.~S9).

We now discuss the origin of this modulation. The local spectral function of the electromagnons alone cannot give a modulation with the period of the spin spiral. For a non-magnetic tip, time-reversal symmetry only allows harmonics of $2\textbf{q}$. For the spin model with single-ion anisotropy, which keeps all the sites of the cycloid equivalent, the local spectral function is uniform (SM Section S6). The experimentally observed modulation requires a tunneling process linear in the local spin, which arises when the tunneling electrons couple to the spins through spin-orbit interaction \cite{Balatsky2002}. It is worth noting that this mechanism is analogous to the spin Hall effect \cite{Dyakonov1971,Sinova2015}, as the surface breaks the out-of-plane mirror symmetry and the electrons spreading from the tip acquire a spin polarization perpendicular to their direction of motion. For a current spreading from the tip into the layer, this coupling gives a term $\eta_i$ that probes the curl of the spin texture around site $i$ (See Methods). For the in-plane cycloid of NiI$_2$, $\eta_i$ is a pure harmonic with the period of the spin spiral, locked to the polarization stripes, with no component at $2\textbf{q}$ (Fig.~\ref{Figure4}d). 
Reversing the bias reverses the direction of the electron flow, and with it the spin acquired by the electrons, so that the contribution of $\eta_i$ to the conductance changes sign. The result is a stripe at $\textbf{q}$ whose contrast is opposite at the two bias polarities (Fig.~\ref{Figure4}d, drawn for $V>0$).

Fig.~\ref{Figure4}e shows the measured $q$ modulation as a function of bias together with the electromagnon density of states $S(E)$. The curve is projected on the stripe phase of the conductance ratio $\ln[G(-V)/G(+V)]$, and thus shows the part odd in bias, which dominates between 2.5 and 6 mV (SM Section S7). The modulation is weak inside the electromagnon gap, grows across the gap edge with opposite signs at the two bias polarities, as expected for this spin-orbit contribution, and is largest at $|V-V_0|\approx4.5$--5 mV, just above the main peak of the electromagnon density of states. At higher bias it decreases and almost vanishes above about 12 mV. This contribution is forbidden in linear response and thus sets in only at finite bias with the inelastic excitations. A quantitative description of this bias dependence requires a microscopic model of the spin-orbit-assisted inelastic tunneling, which is beyond the scope of this work. Summarizing, Fig.~\ref{Figure4} provides an additional, real-space check that the low-energy excitations of Fig.~\ref{Figure3} are related to the electromagnons, since the modulation appears at electromagnon energies, has the period of the spin spiral and is predominantly odd in bias. This provides a methodology to probe and characterize electromagnons at the atomic scale.

\section*{Discussion}

In this work, we have demonstrated the atomic-scale visualization and spectroscopic characterization of electromagnons in a monolayer multiferroic. Using a combination of low-temperature STM and IETS, we probed the spatial and energetic fingerprints of the collective bosonic excitations emerging from the multiferroic phase of monolayer NiI$_2$.
Temperature-dependent STM imaging established the temperature at which the multiferroic stripes melt and, through Monte Carlo simulations of the polarization stripes, set the energy scale for these excitations. Spectroscopic IETS measurements revealed in-gap features at low temperature and a gap of about 3 meV in the excitation spectrum, which we identified as the electromagnon gap opened by the magnetoelectric coupling by comparing with theoretical calculations. Importantly, real-space IETS maps uncovered a periodic modulation with the period of the spin spiral, which appears at electromagnon energies and is predominantly odd in bias, as expected for spin-orbit-assisted inelastic tunneling, providing an additional, real-space check that these excitations are electromagnons.

Our results provide direct evidence of the internal structure of electromagnons and provide a methodology for detecting and analyzing such collective bosonic excitations in two-dimensional systems. This approach establishes a new direction in the exploration of magnetoelectric quasiparticles in van der Waals materials and paves the way for electrically-controlled spintronics and low-energy-consumption technologies.

\section*{Materials and Methods}

\subsection*{Effective spin model for monolayer NiI$_2$}

We have performed theoretical calculations of spin ground states and electromagnon excitations relying on an effective spin model approach. The magnetic order of multiferroic NiI$_2$ can be captured by a $J_1-J_3$ spin model with ferromagnetic first-neighbour and antiferromagnetic third-neighbour exchange interactions and single-ion anisotropy $A_z$ along the out-of-plane direction, so that the system features an easy-plane $xy$ ordering \cite{Fumega2022,Li2023_realisticspinmodelnii2,PhysRevB.106.035156}. The spin Hamiltonian reads

\begin{equation}
    H=\sum_{\langle i,j\rangle} J_1 \textbf{S}_i \cdot \textbf{S}_j + J_3 \sum_{\langle\langle\langle i,j\rangle\rangle\rangle} \textbf{S}_i\cdot\textbf{S}_j+A_z\sum_i (S_i^z)^2,
    \label{eq:Spin_Hamiltonian}
\end{equation}

where the single and triple brackets restrict the indices $i$ and $j$ to first and third neighbor interactions, respectively. All of the above terms can systematically be grouped into diagonal exchange interaction and single-ion anisotropy tensors such that the Hamiltonian can more compactly be written as $H=\sum_{ij} \textbf{S}_i\cdot\overline{\overline{\mathcal{J}}}_{ij}\cdot\textbf{S}_j + \sum_i\textbf{S}_i\cdot\overline{\overline{\mathcal{A}}}_i\cdot\textbf{S}_i$.
We use $S=1$, $J_1=-3.79$ meV and $J_3=1.15$ meV, obtained from the temperature dependence of the stripes (Fig.~\ref{Figure2} and SM Section S5), with $J_3/J_1=-0.3025$ giving a spin spiral of period $9a$ along $\Gamma$--K, and $A_z=0.05|J_1|=0.19$ meV (SM Sections S5 and S6).

This Hamiltonian can be used to capture both the ground state and magnonic excitations of the spin spiral multiferroic. The ground state spin spiral is obtained by minimizing the classical energy of the spin configuration treating each spin operator as a local classical magnetization vector $\textbf{M}_i$. This minimization of the total energy is achieved using the Broyden–Fletcher–Goldfarb–Shanno (BFGS) algorithm. 
To ensure robustness against local minima, the initial parameters are randomized, and the minimization is performed on the order of 200 independent runs. The configuration with the lowest energy among these trials is selected as the final ground state. We verify that independent runs with different random initializations consistently converge to equivalent ground state configurations, confirming the stability of the result.

On top of the ground state, magnonic excitations can be computed using a Holstein-Primakoff-Bogoliubov-de-Gennes (HPBdG) approach. First, all ground state spins are rotated under the action of a unitary matrix $\textbf{U}_{i(jk)}$ such that they point in the $z-$direction \cite{PhysRev.139.A450, Toth2015, Cong2024}. The first index corresponds to the site $i$ within a real-space $N-$site supercell, and the two indices in parenthesis $(jk)$ to the spatial indices of the matrix.
After this rotation, one can map between spin operators and Holstein-Primakoff bosonic creation ($a_i^\dagger$) and annihilation ($a_i$) operators such that along the ground state magnetization direction, we have $S^\zeta_i =  S-a_i^\dagger a_i$, and along an adequately chosen rotating frame the spin operators read $S^\xi_i \approx\sqrt{S/2}(a_i+a_i^\dagger)$ and $S^\eta_i\approx -i\sqrt{S/2}(a_i-a_i^\dagger)$, which corresponds to $S^+_i\approx\sqrt{2S}\,a_i$. In terms of these operators, the rotation under $\textbf{U}_i$ corresponds to a Bogoliubov transformation. By performing these replacements and moving to a momentum space description via a Fourier transform of the creation and annihilation operators, we obtain the linearized spin-wave theory by retaining second order terms in the magnon operators. The magnon operators are then grouped into vectors $\textbf{x}(\textbf{k})=(a_1^\dagger(\textbf{k}),\cdots,a_N^\dagger(\textbf{k}), a_1(\textbf{k}),\cdots,a_N(\textbf{k}))^T$ such that  the HPBdG Hamiltonian can be written as $H_{BdG} = \sum_\textbf{k}\textbf{x}^\dagger(\textbf{k}) H_{BdG}(\textbf{k}) \textbf{x}(\textbf{k})$. This Hamiltonian has the structure \cite{Toth2015}

\begin{equation}
    H_{BdG}(\textbf{k})=
\begin{pmatrix}
A(\textbf{k}) - C(\textbf{k}) & B(\textbf{k}) \\
B^\dagger(\textbf{k}) & A^\dagger(-\textbf{k})  - C(\textbf{k})
\end{pmatrix},
\end{equation}
where each entry $A, B$ and $C$ corresponds to an $N\times N$ matrix with entries

\begin{align}
    A_{ij}(\textbf{k}) &=\left(\textbf{u}_i\cdot\left[\overline{\overline{\mathcal{J}}}_{ij}(\textbf{k})+\delta_{ij}\overline{\overline{\mathcal{A}}}_{i}\right]\cdot\textbf{u}_j^*\right)/2, \\
    B_{ij}(\textbf{k}) &=\left(\textbf{u}_i\cdot\left[\overline{\overline{\mathcal{J}}}_{ij}(\textbf{k})+\delta_{ij}\overline{\overline{\mathcal{A}}}_{i}\right]\cdot\textbf{u}_j\right)/2,  \\
    C_{ij}(\textbf{k}) &=\delta_{ij}\sum_l\left(\textbf{v}_i\cdot\overline{\overline{\mathcal{J}}}_{i}(\textbf{0})\cdot\textbf{v}_l\right)/2. 
\end{align}
With the vectors $\textbf{u}_{i}=\textbf{U}_{i(jx)}+i\textbf{U}_{i(jy)}$ and $\textbf{v}_{i}=\textbf{U}_{i(jz)}$ one can construct the linearized HPBdG Hamiltonian. 
Diagonalizing the BdG Hamiltonian results in a doubled spectrum of positive and negative energy modes $\boldsymbol{\Omega}_i$. The real energies of the magnonic modes can be obtained by multiplying $E_i=\sigma_z \boldsymbol{\Omega}_i$, where $\sigma_z$ acts with +1 (-1) in the positive (negative) energy sector.

The electromagnon eigenmodes arise by including the energy term associated with fluctuations in electric polarization within the spin Hamiltonian. The polarization generated via the inverse DMI is given by $\textbf{P}_i=\lambda \textbf{S}_i\times\nabla\times \textbf{S}_i$, or, in its real-space form, by $\textbf{P}_i = \lambda/6a \sum_j  \textbf{S}_i \times \textbf{r}_{ij}\times \left(\textbf{S}_j-\textbf{S}_i\right)$. This local form averages over the spiral to the uniform macroscopic polarization along $\hat{\textbf{q}}\times\textbf{e}$ of the spin-current mechanism, whereas its modulation at $2\textbf{q}$ gives rise to the polarization stripes. In the spin-spiral ground state, the static polarization $\textbf{P}^0_i$ at each site is perpendicular to the local magnetization $\textbf{M}_i$. We describe the energy cost of the spin fluctuations that modify the local polarization with an effective magnetoelectric coupling
\begin{equation}
    H_D=D\sum_i\left(\textbf{S}_i\cdot\hat{\textbf{P}}^0_i\right)^2,
    \label{eq:HD}
\end{equation}
where $\hat{\textbf{P}}^0_i=\textbf{P}^0_i/|\textbf{P}^0_i|$ is the direction of the static polarization at site $i$. This term is quadratic in the spin operators and therefore enters the linear spin-wave theory directly as a site-dependent correction to the single-ion anisotropy tensor, $\overline{\overline{\mathcal{A}}}_i\rightarrow\overline{\overline{\mathcal{A}}}_i+D\,\hat{\textbf{P}}^0_i\otimes\hat{\textbf{P}}^0_i$, where $\otimes$ is the dyadic product. Since $\hat{\textbf{P}}^0_i\perp\textbf{M}_i$, $H_D$ does not modify the ground state obtained from Eq.~\ref{eq:Spin_Hamiltonian} and only raises the energy of the spin precession towards $\hat{\textbf{P}}^0_i$, which gaps the phason and the tilt modes of the spin spiral (Fig.~\ref{Figure1}c,d). We use $D=3.0|J_1|=11.4$ meV, fixed by the gap of the inelastic spectrum (Fig.~\ref{Figure3}). Semiclassically, these excitations can be interpreted as carrying both fluctuations in magnetization $\textbf{m}_i$ and in polarization $\textbf{p}_i = \lambda/6a\sum_j \textbf{m}_i\times\textbf{r}_{ij} \times(\textbf{m}_j-\textbf{m}_i)$, and therefore correspond to electromagnons.

\subsection*{Monte Carlo simulations}

We computed the temperature dependence of the multiferroic order for monolayer NiI$_2$ using classical Monte Carlo simulations and the spin model for monolayer NiI$_2$ of Eq.~\ref{eq:Spin_Hamiltonian}, on $216\times216$ lattices with periodic boundary conditions. For every spin configuration we evaluate the emergent polarization $\textbf{P}_i$, which gives the polarization structure factor $S_P(\textbf{Q})$ and the amplitude of the polarization stripes $S_P(2\textbf{q})$, as well as the spin structure factor $S(\textbf{q})$ and the uniform polarization. The energy scale of the exchange interactions is obtained by fitting $S_P(2\textbf{q})$ to the measured stripe amplitude (Fig.~\ref{Figure2}g). The details are given in SM Section S5.

\subsection*{Inelastic spectra and spin-orbit tunneling}

The inelastic spectra are compared with the local transverse spin spectral function of the electromagnons, $S(E)=\frac{1}{N}\sum_{i}\sum_{n,\mathbf{k}}w_{i,n}(\mathbf{k})\,\delta(E-\varepsilon_n(\mathbf{k}))$, where $w_{i,n}(\mathbf{k})$ is the weight of the transverse spin fluctuations at site $i$ in mode $n$, computed on a fine grid of $\mathbf{k}$-points of the $9a\times\sqrt{3}a$ supercell. For a non-magnetic tip, the inelastic contribution to \ddIdV at $E=e|V-V_0|$ is proportional to $S(E)$. The spectrum is convolved with a Gaussian, and only its amplitude and broadening are fitted to the group-averaged \ddIdV between 2.0 and 4.5 meV (Fig.~\ref{Figure3}e), and the \dIdV step of Fig.~\ref{Figure3}d is the integral of the same spectrum. The phonon contribution of Fig.~\ref{Figure3}f is the DFT phonon density of states of the multiferroic phase restricted to the out-of-plane, near-$\Gamma$ A$_{1g}$ band, whose amplitude is fitted between 5 and 17 meV.

For the spin-orbit-assisted tunneling of Fig.~\ref{Figure4}d \cite{Balatsky2002}, a current flowing radially from the tip into the layer couples to the spins as $\eta_i=\sum_j(\hat{\textbf{z}}\times\textbf{e}_{ij})\cdot\textbf{S}_j$, where the sum runs over the six nearest neighbors of site $i$ and $\textbf{e}_{ij}$ are the bond unit vectors. A tip at position $\textbf{r}$ with a current spreading $w(\rho)=e^{-\rho/\ell}$, $\ell=2a$, gives the map $C(\textbf{r})=\sum_i w(\rho_i)\left[\hat{\boldsymbol{\rho}}_i\times\textbf{S}_i\right]_z$, 
with $\boldsymbol{\rho}_i=\textbf{r}_i-\textbf{r}$, which for the in-plane cycloid has the period of the spin spiral. The contribution changes sign with the bias polarity. Fig.~\ref{Figure4}e shows the measured $q$ modulation with the electromagnon density of states $S(E)$ only, and no model of its bias dependence is fitted. The measured $q$ modulation, the line spectra and the Fourier weight at higher bias are described in SM Section S7.

\subsection*{Ab initio calculations and harmonic phonons in the multiferroic phase} 

We have performed \textit{ab initio} electronic structure calculations based on density functional theory \cite{HK,KS} (DFT) in monolayer NiI$_2$. Calculations were carried out with the all-electron full-potential linearized augmented-plane-wave method, using a fully non-collinear formalism with spin-orbit coupling (SOC) as implemented in Elk \cite{Elk}. We have used the generalized gradient approximation in the Perdew-Burke-Ernzerhof scheme (GGA-PBE) for the exchange-correlation functional \cite{PhysRevLett.77.3865}. 
The results presented are converged with respect to all the parameters, considering the $9a\times\sqrt{3}a$ supercell with the spin spiral, a $3\times 8 \times 1$ k-mesh, and a vacuum spacing of 20 \AA. 
Before computing the harmonic phonon spectrum, relaxations of the atomic positions in the spin spiral state including SOC require careful convergence of the forces ($10^{-4}$ a.u) and the Kohn-Sham potential ($10^{-7}$ a.u). This step is essential, as the emergent polarization in the multiferroic phase induces a ferroelectric response in the lattice, lowering the crystalline symmetry from space group 164 (P-3m1) at high temperature to space group 8 (Cm) in the low-temperature multiferroic phase.
Harmonic phonon calculations were performed on the Cm multiferroic phase using the real-space supercell approach as implemented in the phonopy code \cite{phonopy}.
The calculations include the spin spiral, SOC and $9a\times\sqrt{3}a$ supercells were used to compute the force sets and dynamical matrix. Finally, $30\times 80 \times 1$ q-meshes were used in the calculations of the phonon DOS and local spectral function.

\subsection*{Sample preparation} 

NiI$_2$ was grown by molecular beam epitaxy (MBE) on highly oriented pyrolytic graphite (HOPG) under ultra-high vacuum conditions (UHV, base pressure $\sim1\times10^{-10}$ mbar). HOPG crystal was cleaved and subsequently out-gassed at $\sim300^\circ$C. High-purity Ni was evaporated from an electron-beam evaporator. Before growth, the flux of Ni was calibrated on an Au(111) at $\sim1$ monolayer per hour. Iodine was evaporated from a Knudsen cell using NiI$_2$ powder as an iodine source (NiI$_2$ decomposes at a temperature of around $\sim400^\circ$C). The sample was grown in an iodine background pressure of $\sim9\times10^{-8}$ mbar and the growth duration was 30 minutes. 
Before the growth, HOPG substrate temperature was stabilized at $\sim100^\circ$C. 

\subsection*{STM measurements}

After the preparation, the sample was inserted into the low-temperature STM (Unisoku LT-STM) connected to the same UHV system, and subsequent experiments were performed at $T = 350$ mK unless otherwise stated. STM images were taken in the constant-current mode unless otherwise stated. d$I$/d$V$ spectra were recorded by standard lock-in detection while sweeping the sample bias in an open feedback loop configuration, with a peak-to-peak bias modulation of 0.25 mV for IETS short-range spectra and 10mV for long-range spectra at a frequency of 757 Hz. 

\section*{Acknowledgments}

This research made use of the Aalto Nanomicroscopy Center (Aalto NMC) facilities and was supported by the European Research Council (ERC-2023-AdG GETREAL (no.~101142364), ERC-2024-CoG ULTRATWISTROICS (no.~101170477), and ERC-2021-StG TITAN (no.~101039500) and the Research Council of Finland (Academy Research Fellow nos.~369367, 368478, 338478, 371757, the Finnish Quantum Flagship project no.~358877, and the Finnish Centre of Excellence in Quantum Materials QMAT no.~374166).  Computing resources from the Aalto Science-IT project and CSC Helsinki are gratefully acknowledged.  

\section*{Competing Interests}
The authors declare no conflict of interest.

\section*{Data availability}

All the data supporting the findings are available from the corresponding authors upon reasonable request.

\section*{Author contributions}
M.A., T.V.C.A., S.K., J.L.L., A.O.F., and P.L. conceived the experiment. M.A., L.J., Z.W., A.K., R.D., and S.K. carried out the sample growth and the low-temperature STM experiments. M.A., T.V.C.A., Z.W., and A.O.F. analysed the STM data. T.V.C.A., J.L.L., and A.O.F. developed the theoretical model. M.A., T.V.C.A., A.O.F., and P.L. wrote the manuscript with input from all co-authors.

\bibliography{NiI2}

\end{document}